\documentclass[copyright,creativecommons,noderivs,noncommercial]{eptcs}
\usepackage[latin9]{inputenc}
\usepackage{units}
\usepackage{amstext}
\usepackage{graphicx}
\usepackage[sort]{cite}

\makeatletter
\providecommand{\event}{PC 2016} % Name of the event you are submitting to
\usepackage{breakurl}% Not needed if you use pdflatex only.
\usepackage{hyperref}

\title{Quantum Probability as an Application\\ of Data Compression Principles}
\author{Allan F. Randall
\institute{Dept. of Information \& Communication Eng. Tech.\\
Centennial College\\  %\thanks{A fine university.}\\
Toronto, Ontario, Canada}
\email{research@allanrandall.ca}
}
\def\titlerunning{Quantum Probability as an Application of Data Compression Principles}
\def\authorrunning{Allan F. Randall}

\makeatother

\begin{document}
\maketitle
\begin{abstract}
Realist, no-collapse interpretations of quantum mechanics, such as
Everett's, face the probability problem: how to justify the norm-squared
(Born) rule from the wavefunction alone. While any basis-independent
measure can \emph{only} be norm-squared (due to the Gleason-Busch
Theorem) this fact conflicts with various popular, non-wavefunction-based
phenomenological measures---such as observer, outcome or world counting---that
are frequently demanded of Everettians. These alternatives conflict,
however, with the wavefunction realism upon which Everett's approach
rests, which seems to call for an objective, basis-independent measure
based only on wavefunction amplitudes. The ability of quantum probabilities
to destructively interfere with each other, however, makes it difficult
to see how probabilities can be derived solely from amplitudes in
an intuitively appealing way. I argue that the use of algorithmic
probability can solve this problem, since the objective, single-case
probability measure that wavefunction realism demands is exactly what
algorithmic information theory was designed to provide. The result
is an intuitive account of complex-valued amplitudes, as coefficients
in an optimal lossy data compression, such that changes in algorithmic
information content (entropy deltas) are associated with phenomenal
transitions.
\end{abstract}

\section{Introduction}

Approaches to quantum foundations based on Everett's no-collapse and/or
wavefunction realism postulates \cite{Everett:1957tb} inevitably
hit the probability problem \cite{Hartle:1968vo,Graham:1973vu,Farhi:1989up,Kent:1990we,Squires:1990wf,Gutmann:1995te,Deutsch:1999ea,Neumaier:1999te,Zurek:2004bw,Albert:2010tm,Wallace:2010tg}:
how to justify the norm-amplitude squared (Born) rule \cite{Born:1926ul}
from the wavefunction alone. While any probability based on an objective,
basis-independent measure can \emph{only} be norm-squared, due to
the Gleason-Busch Theorem \cite{Gleason:1957vz,Busch:1999hp}, this
is often seen as being in conflict with various intuitive and popular
ideas and conceptions of probability. Most particularly, it is felt
that some kind of phenomenological counting measure, rather than a
wavefunction-based measure, ought to be employed, such as observer,
outcome or world counting \cite{Everett:1957tb,Graham:1973vu}. While
little rigorous argument has been presented for this feeling, it appears
to be quite strong, to the extent that such a non-wavefunction-based
measure is widely cited as a requirement for any pure wavefunction-based
(no-collapse) interpretation of quantum probability.

This would seem, on the face of it, a somewhat contradictory demand:
since wavefunction realism is the hall-mark of no-collapse interpretations,
why are we requiring anything but a purely wavefunction-based measure
of them? The answer may lie in the lack of an intuitive explanation
for any pure wavefunction measure---how can wavefunction amplitudes
actually ``count'' towards a probability measure, in the way marbles
count when we pick one at random from a bag? Instead of something
straightforward, we have counterintuitive puzzles like complex counts
and negative probabilities. What is needed, then, is \emph{not} a
Born rule proof, which we already have (actually we have lots of them!).
What is needed is an intuitive story to tell about why complex-valued
counts can make sense, connecting the ontological wavefunction to
the phenomena without appealing to any reality beyond the wavefunction
and without making phenomenal entities the basis of the probability
counts.

I will explore the feasibility of an algorithmic probability measure,
by making the minimal assumptions about such a measure as are consistent
with \emph{(1) }wavefunction realism and \emph{(2) }the assumption
that quantum probability should be algorithmic and information-theoretic
in nature. The result is a derivation of the Born rule that provides
an intuitive rationale for complex counts, conceiving them as changes
in algorithmic information content (entropy deltas) associated with
phenomenal transitions. It remains true to the realist, objectivist
intent of no-collapse interpretations, and gives an intuitive story
for the role of probability interference under such interpretations.
Although only the impact on the Born rule is discussed here, this
is intended as a first step towards a complete and consistent algorithmic
interpretation of quantum mechanics (see also \cite{Randall:2013wb}).

\section{The Probability Objection}

We assume with Everett the doctrine of wavefunction realism \cite{Everett:1957tb}:
\begin{quote}
``A wave function that obeys a linear wave equation everywhere and
at all times supplies a complete mathematical model for every isolated
physical system without exception.''
\end{quote}
It might seem that this makes probabilities meaningless since all
possible outcomes are realized. However, one can model observers within
the wavefunction as machines or robots, with sensors to collect information,
circuitry to process that information, and memory to store the information
so that it may affect future behaviour. We then seek to calculate,
given a phenomenon being experienced by one of these observers, what
the transition probabilities are for all subsequent possible ``continuer''
experiences for that observer, as described by the unitary evolution
of the wavefunction.

Unfortunately, it is not at all obvious how one is supposed to go
about performing such a calculation. One problem is that the very
idea of multiple different mutually incompatible experiences is alien
to our common experience. We do not wake up one morning to find five
other copies of ourselves running around the house. However, a well-known
and purely classical thought experiment can yield just this scenario
without requiring any quantum mechanics. Imagine an ingenious replicator
machine that can make a complete (or complete-enough) scan of the physical
state of a human being (including the brain) and then build any number of
perfect working replicas or copies of that human being. Assume for convenience
that the original body and brain are destroyed in the scanning process,
so there is no temptation to confer a privileged status on any one
resulting copy.

\begin{figure}
\noindent \begin{centering}
\includegraphics[scale=0.4]{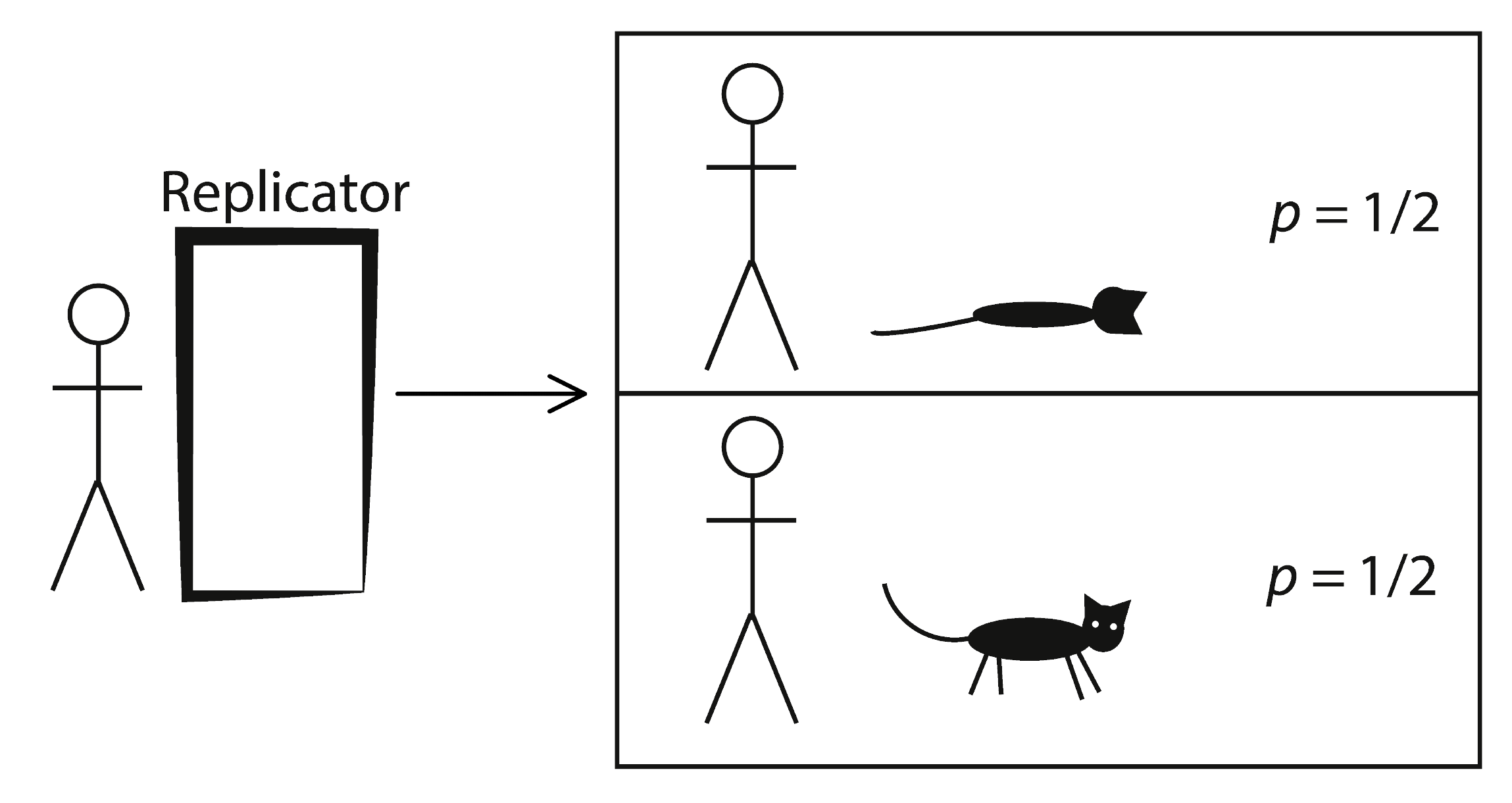}
\par\end{centering}

\caption{Replicator copies: \emph{two} copies yields \emph{two} counts of equal
measure.}
\end{figure}

Imagine you are about to be so copied. You are told that the scanner
produces two identical copies, each of which exits through a separate
exit door. Each door goes to a small room with
a closed box on the floor. The rooms are identical, except that in
one room, the closed box has a live cat inside, while in the other
room, it contains a dead cat. You are told that when your copies exit
the machine, they will be asked to open the box that appears in front
of them. What, you are asked, is the probability, from \emph{your
}perspective, of finding a live cat in the box? Obviously, the answer
is $\nicefrac{1}{2}$, since there are two copies or continuers made,
and you could find yourself equally well as either continuer after
replication.

Probability objectors to Everett reason that wavefunction probabilities
should work the same way. In the original cat experiment \cite{Schrodinger:1980tw},
the wavefunction computed two and only two continuers, and hence
an \emph{a priori} probability of $\nicefrac{1}{2}$ for each. Or,
more generally, for $n$ continuers/copies, we have
\begin{equation}
p\left(m_{i}\mid m\right)=\frac{1}{n},
\end{equation}
resulting in a simple, flat probability distribution across all the
Everettian ``branches'' (or ``worlds'' or ``copies'' or ``continuers'').
Of course, we know this is \emph{not }the actual probability rule,
which is instead Born's norm-amplitude-squared rule.

To avoid the reification of subsystems, which goes against wavefunction
realism, we will express the wavefunction in Everettian terms, by
decomposing the entire wavefunction of the universe according to an
observation basis that is defined relative to an observer experience
$m$ and its corresponding continuers:
\begin{equation}
\left|\psi\right\rangle =a_{1}\left|\psi_{1}\right\rangle +\cdots+a_{n}\left|\psi_{n}\right\rangle,
\end{equation}
so that, according to the Born rule,
\begin{equation}
p\left(m_{i}\mid m\right)=p\left(\psi_{i}\mid\psi\right)\propto\left|\left\langle \psi_{i}\mid\psi\right\rangle \right|^{2}=\left|a_{i}\right|^{2}.
\end{equation}

This formulation makes a couple of assumptions:
\begin{enumerate}
\item It is possible to treat each continuer as an independent, orthogonal
basis element.
\item It is possible to treat the wavefunction $\left|\psi\right\rangle $
as a linear combination of these continuers.
\end{enumerate}
Neither of these assumptions is ultimately problematic, however. If
we suspend assumption 1, it means there are at least two continuers
that are not linearly independent, and so would have to be included
as multiple copies \emph{within one Everettian branch}. Since we,
in fact, do not generally experience multiple continuers within our
branch or ``world''---since nothing like replicator machines seem
to exist---assumption 1 does not present a serious problem. We could
easily suspend it, by using the Born rule to assign a probability
to each $\left|\psi_{i}\right\rangle $, and then split $\left|\psi_{i}\right\rangle $'s
measure equally amongst its internal continuers or copies.

We could also suspend assumption 2. If the linearly-independent continuers
of $m$ cannot be linearly combined to reproduce $\left|\psi\right\rangle $,
then we simply add a unit-norm remainder state $\left|\psi_{D}\right\rangle $,
in which no internal continuer of $m$ can be found, but which allows
the expansion in terms of continuers to add up to the wavefunction:
\begin{equation}
\left|\psi\right\rangle =a_{1}\left|\psi_{1}\right\rangle +\cdots+a_{n}\left|\psi_{n}\right\rangle +a_{D}\left|\psi_{D}\right\rangle, 
\end{equation}
\smallskip{}
so that $\left|\psi_{D}\right\rangle $ might be called the ``dead
branch'', since it is the world(s) or branch(es) without any continuer
(the original observer having apparently ``died'' in that branch).

In most laboratory applications of Born's rule, however, both assumptions
1 and 2 clearly hold, so we will continue to talk in terms of these
two assumptions, without loss of generality.

Experiment upholds the Born rule, which calculates probabilities based
on amplitudes, so that equal amplitude means equal probability, higher
amplitudes mean higher probability, and lower amplitudes mean lower
probability. In fact, Everett \cite{Everett:1957tb} proves that so
long as one assumes ``amplitude dependence''---that the measure
of an outcome is a function solely of amplitude---then the Born rules
follows.

The point of the objectors to Everett is that there is no \emph{a
priori }justification for amplitude-dependence under the assumptions
of wavefunction realism. A lower-amplitude branch is not somehow
``less computed'' by the wavefunction than the high-amplitude branches.
So how do we justify giving it a lower measure, in the way we justified
counting copies in the replicator experiment? Thus far, it seems that
the world or outcome counters have the upper hand, since counting
outcomes (or worlds or continuers) would seem to be the equivalent
of counting replicator copies. However, as we will soon see, the situation
is not as simple as that.

\section{Response to the Probability Objection}

The most mathematically straightforward, and most formalistic, response
to this objection (although not one very commonly used) is simply
to point to the Gleason-Busch theorem \cite{Gleason:1957vz,Busch:1999hp},
which tells us that Born's rule is mathematically the \emph{only }possible
probability rule, under an assumption of additivity for the probabilities
(demanded by the nature of probabilities) along with the assumption
of ``noncontextuality'' or ``basis-independence'', which states
simply that the measure used to compute probability is independent
of the basis used to expand $\left|\psi\right\rangle $. Thus \emph{any}
basis $\left\{ \dots,\left|\psi_{i}\right\rangle, \dots\right\} $
that contains continuer state $\left|\psi_{i}\right\rangle $, regardless
of whether the other basis elements correspond to continuers, can
use the \emph{same} probability rule to calculate \emph{the same}
probability for $m_{i}$.

In Everettian terms, noncontextuality is often thought of in terms
of branch-dependence, since this is a consequence of noncontextuality,
and one that can be intuitively applied to branches. Mathematically,
branch-dependence states that there exists a function $f$ such that
\begin{equation}
p\left(m_{i}\mid m\right)\propto f\left(\left|\psi_{i}\right\rangle \right),
\end{equation}
so that the measure of a particular continuer is a function only of
that continuer's branch of the wavefunction (other than for normalization).
From this, one can prove amplitude-dependence, and from there the
Born rule, via either Everett's proof \cite{Everett:1957tb} or (for
dimensions greater than two) via certain versions of Gleason's original
proof \cite{2012arXiv1202.2728L}.

But \emph{is} this assumption at all intuitive, within wavefunction
realism? The probability objectors say no, since the wavefunction
computes all the continuers, and probability is something that can
only be calculated from the perspective of the observer. Hence, we
have no reason to assume that the measure we use must be noncontextual,
since it is intended to yield only the probability from the perspective
of this observer ($m$). This is arguably a subjectivist conception
of the probability measure: all that matters is how many subjectively
indistinguishable states there are. If there are $n$ such states,
then the probability\emph{---from the perspective of $m$}---of finding
oneself in \emph{one }of those $n$ continuer states or copies \emph{must
}be $\nicefrac{1}{n}$. Or so goes the reasoning.

There is, unfortunately, little in the way of rigorous argumentation
for this ``world-counting'' or ``observer-counting'' assumption.
When it is used, it is usually simply assumed. Note that in many very
straightforward applications of probability theory, the categories
used are subjective indistinguishables, while the countables are still
assumed to represent actually existing, objective entities. For instance,
if I pick a marble at random from a bag with 7 blue and 3 red marbles,
there is a 30\% chance of picking ``red''. Now ``red'' is a subjective
category, true. But the actual objective entities being picked are
just marbles. I could distinguish them differently if I chose. I choose
``red'' versus ``blue'' because that makes sense for me. However,
the objects counted are still the actual marbles, with the objective
count of things falling within my subjective category placed in the
numerator, and the total objective count of things in the denominator.
In neither case am I ``counting'' subjective indistinguishables,
but rather objectively existing things. So we still call this an ``objective''
probability rule, not a subjective one (so long as we consider the
marbles themselves to objectively exist).

It might seem that counting distinct observers is like counting copies
in the replicator experiment, but a little reflection reveals otherwise.
Imagine that, instead of \emph{two }copies, there are now \emph{three
}copies, three exit doors and three rooms with closed boxes. One room
has a box with a dead cat, while the other two have a box with a live
cat. What now are the probabilities of seeing a live cat, given the
method of counting copies? It would seem now that the probability
of seeing a live cat is clearly $\nicefrac{2}{3}$, given that there
are three copies and two of them see a living cat.

But are there \emph{really} two live-cat copies, in other words \emph{two
distinct observers} that see a live cat? Probably in a realistic scenario,
the answer is clearly yes, since there will be at least subtle but
real differences in the two rooms with the live cats. However, if
we really can make the rooms \emph{absolutely identical} in every
respect, other than the cat inside the box, then it is arguable that
there is really only one observer that sees a live cat, since there
is no formal difference between the two observers, and under wavefunction
realism, observers are considered to be merely emergent phenomenon
arising from a purely formal computational system. If the \emph{experiences}
of the two live-cat copies are absolutely identical, then surely---from
the perspective of this observer---there is only \emph{one} indistinguishable
experiential continuer-state, and thus these two live-cat rooms actually
constitute a \emph{single }observer.

And clearly, the probability of seeing a live cat remains $\nicefrac{2}{3}$,
in either case, for if we did the same experiment on someone repeatedly,
they would measure a 2:1 ratio of live:dead outcomes, regardless of
how precisely (or imprecisely) we made the two live-cat rooms match. 

In other words, there is \emph{one }continuer/observer present while
there are \emph{two }physical copies/rooms. Thus, we are \emph{not},
after all, counting observers. One of the two observers actually consists
of two \emph{copies}, and \emph{this} is what we count. The number
of copies, then, is comparable to wavefunction amplitude, while the
observer-count (or world-count) is itself more or less irrelevant
to the measure.

\begin{figure}
\noindent \begin{raggedright}
\includegraphics[scale=0.4]{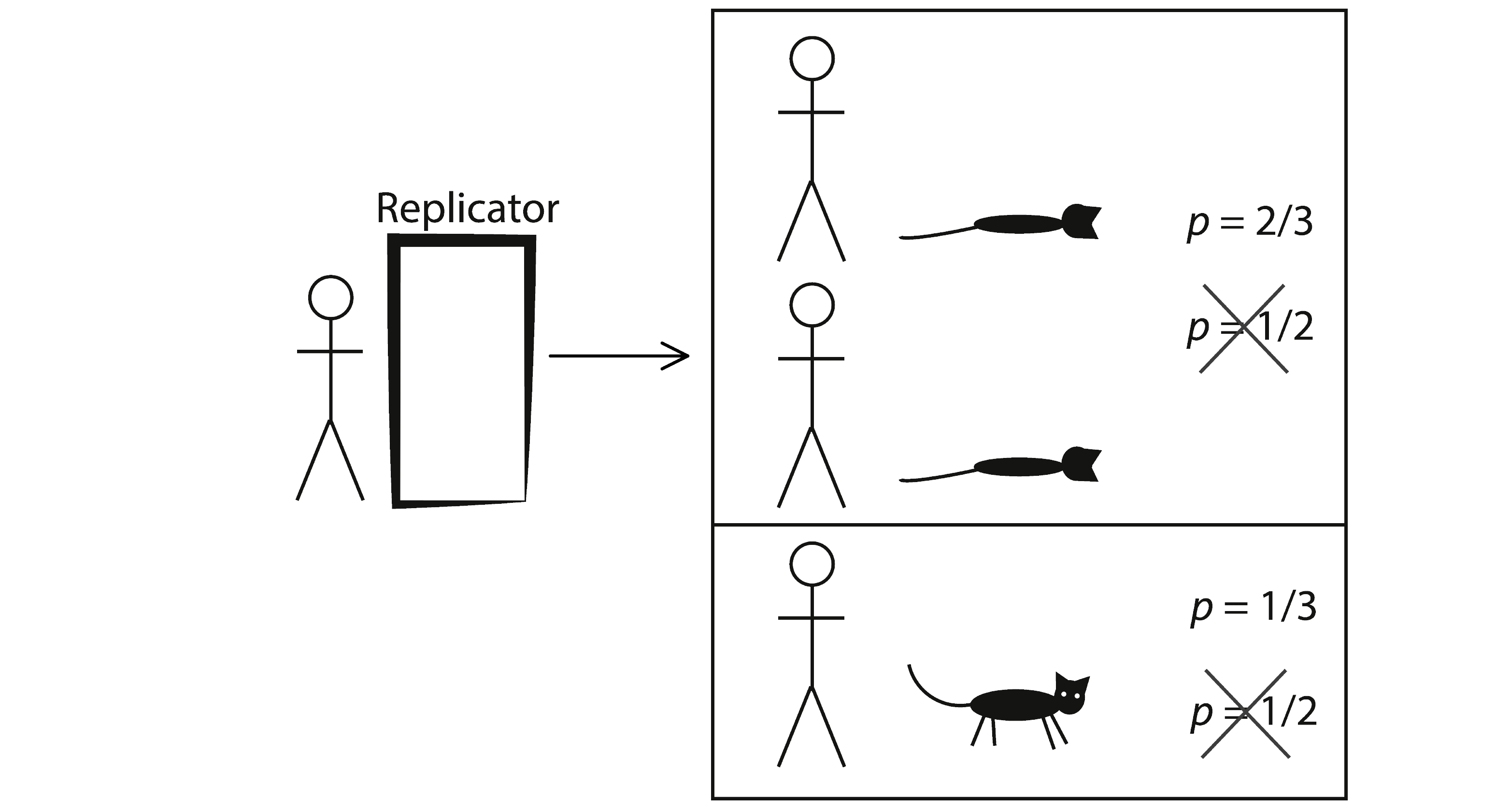}
\par\end{raggedright}

\caption{Replicator copies: \emph{two} observers, \emph{three }copies yields
\emph{three} objective counts of equal measure in \emph{two }subjective
categories.}
\end{figure}

Of course, this by no means proves that amplitudes are\emph{ }the
appropriate countable for wavefunctions. However, if amplitudes could
be counted in a classical manner to produce probabilities, then there
would be a much stronger argument for equating copy-counts to amplitudes.
But amplitudes are complex-valued, which means they can exhibit destructive
interference. This makes the comparison with the replicator rooms
obscure. What would it mean to send someone through a replicator machine,
that does everything required to produce both live-cat copies, but
somehow, because these two copies have opposite phase, they cancel
each other out, and now we have \emph{no one }observing a live cat?
How can copies cancel each other out, or interfere with each other?

We can perhaps make things less obscure by replacing the replicator
copies with multiple copies simulated in a computer, in a virtual
reality. For here we have a situation (a computer program) whose functioning
we can more easily consider purely on its formal properties, as a
Turing machine, since the ``observers'' and ``rooms'' are simulated
and not really constituted from actual physical objects. This is actually
a benefit within the Everettian context, since we are \emph{supposed}
to be calculating probabilities, in the first place, from the formal
structure of the wavefunction alone.

In this scenario, the program is actually computing\emph{ }the live-cat
observer \emph{twice}. So twice the computing cycles, computer memory
and other computing resources are being used to simulate the two observers.
Perhaps, then, they should count as ``two copies'', even though
they are subjectively indistinguishable as observers. However, once
again, the fact that the amplitudes can interfere call this interpretation
into question. For the resources dedicated to the calculation of the
wavefunction are, it would seem, the same whether the amplitude is
high or low. And two interfering high amplitudes can result in \emph{no
}amplitude for a particular outcome. Yet, it is not clear that ``no
resources'' went into the computation of that outcome. In a sense,
it is obvious that no resources went into its computation, since the
wavefunction does not even compute that outcome at all! Yet, if we
change the phases just \emph{slightly} so that there is a \emph{tiny
}remaining amplitude for that outcome, then it is suddenly again very
unclear how we could argue that there is somehow ``very little''
computing resources dedicated to that outcome, just because the amplitude
is low. The fact that the amplitude is low does not, surely, mean
that it takes less resources to compute it, provided that it \emph{is
}in fact computed. So, while a straight simulation of the replicator
example as a computer simulation would seem to strengthen the copy-count/amplitude
analogy, any incorporation of probability interference seems to call
the analogy into question.

\section{The Algorithmic Assumption}

In order to find anything to count within a ``purely formal'' system
like a pure wavefunction, we have to have some idea just \emph{what
}a pure formalism is. While there are competing conceptions of formality,
the idea of computation as the core of all formal systems carries
a great deal of weight. G{\"o}del \cite{Godel1931hj} claimed, for instance,
that Turing's machines provided ``a precise and unquestionably adequate
definition of the general concept of formal system'' and that ``the
term `formal system' or `formalism' should never be used for anything
but this notion.''

I will therefore adopt the minimal \emph{algorithmic assumption }needed
to justify the application of algorithmic probabilities to emergent
observers:
\begin{quote}
We can, if we wish, model an observer's \emph{experience }as a set
of \emph{finitistic algorithms} (Turing machines) each of which corresponds
to an abstract computer program that \emph{sufficiently and uniquely
describes} that observer experience.
\end{quote}
I will refer to any such sufficient and unique algorithmic description
as ``outputting'' that observer/\allowbreak experience. Note that some convention
must be adopted to interpret a program's state in terms of ``output'',
such that a program cannot be said to describe/output an experience
unless it is distinguished from any other experiences the same program
might also incidentally describe.

This terminology is not meant to rule out the possibility that an
experience may only be output ``in the limit'' of an infinite number
of computational steps of a nonhalting machine (see \cite{Randall:2006ww,Putnam:1965tc,Gold:1965uo,Burgin:1983tf}).
Note that in spite of the finitistic nature of computational models,
the continuum can\emph{ }be modelled in such systems. Also, no assumption
is made here that there is either a finite or infinite number of continuers
for any given experience.

\emph{}Under algorithmic probability, two different continuers might
both be computed, but one may still have higher probability, since
it is output by many more machines than the other---just as the replicator
might produce more physical copies that correspond to one continuer
than another. But instead of counting physical copies, as in the replicator
example, in this purely formal ontology, we count \emph{programs}.

To calculate a probability for a given experience $m$, we must find
all the valid continuer experiences $\left\{ m_{1},\dots, m_{n}\right\} $,
$0\leq n\leq\infty$. We then need to find a way of determining, from
the given continued experience, $m$, what the probability $p\left(m_{i}\mid m\right)$
is for some particular valid continuer $m_{i}$. In the quantum context,
the continued experience must be output by the pre-measurement wavefunction
of the universe (under some appropriate interpretation of ``output'').

If we accept the view that formal systems are essentially Turing machines,
or abstract computer programs, then what are the ``countables''
of a purely formal system, in which observers are merely computed
emergent properties? If we take the wavefunction of the universe as
the only existing thing, and treat it as a purely formal system, then
does this not mean that we have just exactly \emph{one }thing to count,
namely the wavefunction of the universe considered as a program?

The answer to this seems to be no, since the wavefunction can, in
fact, be computed by an infinite number of distinct programs. If we
consider each such program code to be a distinct encoding of the wavefunction,
then the wavefunction ``counts'' as far more than a single object---we
must count every distinct program code that generates the wavefunction.
The distinct encodings are all different, but they all encode the
same algorithm for generating our wavefunction. In a sense, they are
different machines that perform the same calculation, just as I can
add 2+2 on an abacus or a pocket calculator.

Solomonoff \cite{Solomonoff:1960via,Solomonoff:1964t1,Solomonoff:1964t2}
initiated the field of algorithmic probability with the discovery
of a well-defined algorithmic measure of information. The ``information''
or ``entropy'' $H\left(x\right)$ of a code or symbol sequence $x$
can be defined more or less as the size or length in bits of $\mathcal{C}\left(x\right)$,
the ``optimal compression'' of $x$, meaning the shortest self-delimiting
program that generates $x$:
\begin{equation}
H\left(x\right)=\left|\mathcal{C}\left(x\right)\right|.
\end{equation}

A program is ``self-delimiting'' if one can tell from the encoding
itself where the end of the program is. The probability of a randomly
chosen Turing machine (or program) generating $x$ is then simply
\begin{equation}
p\left(x\right)=2^{-H\left(x\right)}.
\end{equation}

A program $x$ is considered to be ``algorithmically random'' if
and only if it is incompressible:
\begin{equation}
x=\mathcal{C}\left(x\right).
\end{equation}

So long as you can further compress $x$, it must contain some kind
of pattern or symmetry, else you would not have been able to shorten
your description of it. 

The strict expression of $p\left(x\right)$ is more complicated than
indicated above, and involves \emph{summing} probabilities over all
the infinity of programs that generate $x$, rather than simply taking
the shortest program. However Solomonoff \cite{Solomonoff:1978ez}
proved that summing over \emph{all} programs (even an infinite number
of them) was \emph{almost} the same thing as taking the shortest program,
since the shorter programs contribute exponentially more to the probability
than the longer programs. The result is a semi-measure rather than
a measure, due to the fact that some encoding are nonhalting and produce
no output, but Solomonoff converts this to a proper measure by normalizing.

One potential problem with the algorithmic measure is the question
of its objectivity. I have thus far assumed that we have adopted a
particular computational language in which to encode our programs,
yielding a particular bit-count for any program. However, choice of
a different language will yield different bit counts and hence a different
probability measure. Nonetheless, Solomonoff's invariance theorem
\cite{Solomonoff:1978ez} proves that the measure is invariant between
languages, up to an additive constant given by the size of the translation
manual between the languages. Thus, for simple languages, it does
seem that the measure should be very close to an objective measure.

I cannot \emph{prove} that algorithmic probability is the correct
\emph{a priori} measure for continuers, of course, since we cannot
(at least not currently) completely formalize observers and their
method of interfacing with the rest of the wavefunction. Nonetheless,
we will take the algorithmic information-theoretic perspective as
a fundamental assumption of this paper. If we can show that such a
measure is consistent with the wavefunction and with the Born rule,
with an intuitive \emph{a priori }justification akin to our count
of replicator copies, then algorithmic probability will make for a
very intuitive choice for an \emph{a priori }measure. At the very
least, we can expect it to give world-counting a run for its money!

\section{Transition Probabilities\label{sub:Continuer-Entropy-1}}

How do we calculate an algorithmic probability for continuer $m_{i}$
given pre-measurement observer $m$? What we cannot do is to pick
$m$ apart, in order to count the number of bits in some subsystem
of $m$ associated with the observer's experience, as this would reify
subsystems, and we are looking for a purely formal probability rule.
This leaves us with $m$ in its entirety for which to calculate an
entropy.

Presumably, a continuer state $m_{i}$ and its continued state $m$
may or may not share a lot of mutual information. If they do, then
$m_{i}$ can be described in very few bits once we already have a
description of $m$. Otherwise, $m_{i}$ will require many bits to
describe, even given a description of $m$. Those continuers that
share very little information will have very low probability, since
it will take a large number of bits to describe them (we will call
these ``maverick continuers''). Continuers $m_{i}$ that have significant
probability will be those with high mutual information content with
the continued state $m$.

Thus, the unshared residual information required to describe a continuer
$m_{i}$ given $m$ is the change in entropy (or information) in transition
from $m$ to $m_{i}$, meaning the size of the optimal compression
of $m_{i}$ given that we already have $m$:
\begin{equation}
\triangle H_{i}=H(m_{i}|m)=H(m_{i})-H(m_{i}:m)=H(m_{i})-H(m),
\end{equation}
where $H(x:y)$ is defined as the ``mutual information'' between
$x$ and $y$, or the optimal number of bits it would take to generate
$x$ if already given $y$ ``for free'', so that $H(x:y)=H(x)-H(x|y)$.

Thus, the probability of continuer $m_{i}$ is based on the (typically
small) $\Delta H_{i}$ associated with the transition $m\rightarrow m_{i}$.
We will call $\Delta H_{i}$ the ``alteration'' bits for the transition,
as it is the number of bits that would have to be added to the optimal
compression of $m$ to generate the optimal compression of $m_{i}$.

The Solomonoff probability $p\left(m_{i}\mid m\right)$ of the transition
$m\longrightarrow m_{i}$ is:
\begin{equation}
p(m_{i}|m)=\frac{p\left(m_{i}\right)}{p\left(m\right)}\propto2^{-\triangle H_{i}}=2^{-\left(H(m_{i})-H(m)\right)}=\frac{2^{-H(m_{i})}}{\sum_{k=1}^{n}2^{-H(m_{k})}}.
\end{equation}

This probability rule has \emph{nothing to do with quantum mechanics},
being a general rule for calculating probabilities for multiple continuers
in a formal system, using algorithmic probability. The rule is clearly
different from counting branches or observers to compute probabilities
(which would set $p(m_{i})=\frac{1}{n}$), since the entropies of
the individual branches are not necessarily equal ($\triangle H_{1}\neq\triangle H_{2}$).
Since there exists a function $f$ such that $p(m_{i}|m)\propto f(m_{i})$,
it follows that the unnormalized measure of a branch depends only
on the information content of \emph{that} branch, as the content of
the prior state (or of the other continuers) serves only to normalize
the measure. This is, of course, simply a form of branch-dependence,
so if it were possible to apply this framework to quantum mechanics,
the Born rule would necessarily hold.

\section{Quantum Transition Probabilities\label{sub:Synthetic-Lossy-Data}}

To apply our algorithmic measure to continuer quantum states, we need
to interpret such states as optimal compressions. As it turns out,
it is very natural to interpret a quantum state as an optimal compression.
Established real-world data compression algorithms intended for synthetic
decompression (into something experienced by an observer), are typically
highly lossy, and are usually some form of ``transform'' compressor,
using some variation on the discrete Fourier transform (or DFT):
\begin{equation}
\mathcal{F}\left(\left\{ \left\langle x\mid\psi\right\rangle :x=1\cdots N\right\} ,R\right)=\left\{ \left\langle k\mid\psi\right\rangle :k=1\cdots R\right\},
\end{equation}
quantizing the original $N$ data points $\left\langle x\mid\psi\right\rangle $
into $R\,$ frequency amplitudes $\left\langle k\mid\psi\right\rangle $,
where $R\ll N$, so that the inverse transform
\begin{equation}
\mathcal{F}^{-1}\left(\left\{ \left\langle k\mid\psi\right\rangle :k=1\cdots R\right\} ,N\right)=\left\{ \left\langle x\mid\psi\right\rangle :x=1\cdots N\right\} 
\end{equation}
decompresses back to the original data set at resolution (dimensionality)
$N$, with
\begin{equation}
\left\{ \left|x\right\rangle =\left[e^{i\frac{1}{R}x\tau}\cdots e^{i\frac{k}{R}x\tau}\cdots1\right]:x=1\cdots N\right\} \,\,\,\textrm{and\,\,\,}\left\{ \left|k\right\rangle =\left[e^{-ik\frac{1}{N}\tau}\cdots e^{-ik\frac{x}{N}\tau}\cdots1\right]:k=1\cdots R\right\} 
\end{equation}
being the orthonormal ``spatial'' and ``frequency'' bases, respectively,
and where $\tau$ is angular unity, the circumference of the unit
circle (more precisely, $\tau$ is some appropriate minimum-bit-length
finite-precision approximation to angular unity, since no actual data
compression takes place on a machine with infinite precision). Note
that the Fourier transform is a unitary transformation and is the
solved form of Schr{\"o}dinger's equation. 

We thus interpret $\left|\psi\right\rangle $, when expressed in a
pure frequency basis of dimension $R$, to be the optimal compression
of pre-measurement observer state $m$, so that $H\left(m\right)=bR$,
where $b$ is the number of bits used by the compression algorithm
to represent a single complex amplitude.

The compressed amplitude list itself, being an optimal compression,
is algorithmically random, containing no further structure or symmetry.
Once transformed to the spatial basis, where the dimensionality $N$
is larger, it is no longer an optimal compression, but nonetheless
encodes the same wavefunction, and so has the same entropy or information
content. Since Fourier transforms are unitary, the unitary evolution
of the wavefunction is in the same category: it does not increase
entropy. However, when an observer splits (at least from that observer's
perspective) then bits have been added to the system, and the optimal
compression will be greater than $R$ by an amount we may call $\triangle H_{i}$.

Thus, the algorithmic transition rule from the previous section can
be readily applied to the inner product vector space of Fourier compressions.
Note that the Solomonoff probability for $m_{i}$ is based on a \emph{different}
DFT than the one for $m$, with a higher dimensionality and entropy.
The probability of $m_{i}$ is based on $\Delta H_{i}$, with the
entropy of $m$ serving only to normalize the measure (as in non-quantum
version):
\begin{equation}
p(m_{i}|m)=\frac{p\left(m_{i}\right)}{p\left(m\right)}\propto2^{-\Delta H_{i}}=2^{-\left(\left|\mathcal{F}{}_{min}(m_{i})\right|-\left|\mathcal{F}{}_{min}(m)\right|\right)}=\frac{2^{-\left|\mathcal{F}{}_{min}(m_{i})\right|}}{\sum_{k=1}^{n}2^{-\left|\mathcal{F}{}_{min}(m_{k})\right|}}.
\end{equation}
Thus, as with the non-quantum version, there exists a function $f$
such that $p(m_{i}|m)\propto f(m_{i})$, so that the measure of a
continuer depends only on \emph{that }continuer branch, not on the
context of the other possible branches (other than for normalization).
This is branch-dependence translated to a periodic (DFT) context,
and it rules out world-counting as the basis for a probability rule.

Thus, if we apply a measure on the discrete Fourier spectrum, it must,
for each frequency, depend only on the individual amplitude for that
frequency (other than for normalization). This makes intuitive sense,
since the overall list of amplitudes is an optimal compression, and
hence algorithmically random, so there can be no structure or symmetry
found within it, with which to extract anything further that could
be informative for our measure.

A corollary to branch-dependence is basis-independence (noncontextuality).
If our measure depends \emph{only }on the given branch, then it cannot
depend on the complement of the branch (the remainder of the wavefunction)
yielding ``complement-independence''. And if the complement does
not matter, then it cannot matter what basis we consider $\left|\psi_{i}\right\rangle $
to be a part of. Given this noncontextuality, Gleason-Busch \cite{Gleason:1957vz,Busch:1999hp}
assures us that all measures of projections onto subspaces of the
vector space are norm-squared measures, and the probability of observer
state $m$ in pre-measurement universe $\left|\psi\right\rangle $
becoming continuer $m_{i}$ in post-measurement universe $\left|\psi_{i}\right\rangle $
follows the Born rule:%\emph{
\begin{equation}
p(m_{i}|m)\propto\left|\left\langle \psi_{i}|\psi\right\rangle \right|^{2}=\left|a_{i}\right|^{2}\label{eq:BornRule}.
\end{equation}
%}

\section{Discussion}

We have seen that algorithmic probabilities based on wavefunction
realism lead very naturally to the Born measure. It might seem odd
that norm-amplitude alone can tell us the measure of a continuer,
when algorithmically we are supposed to be using a bit-counting measure,
not an ``amplitude-counting'' measure. Since each amplitude occupies
the same number of bits in a DFT, it might appear that amplitude-counting
cannot possibly be a form of bit-counting. But recall that it is the
$\triangle H_{i}$ alteration bits for continuer $m_{i}$ that are
counted, not bits \emph{within }a DFT representation, which would
violate wavefunction realism. 

Imagine the extreme case of just two continuers, $m_{1}$ and $m_{2}$
(with no non-continuers) where $\mathcal{C}\left(m_{1}\right)$ is
virtually identical to $\mathcal{C}\left(m\right)$, but for a few
minor alterations, while $\mathcal{C}\left(m_{2}\right)$ is radically
different (perhaps a maverick continuer). This could be the case even
if the \emph{apparent} difference between the two continuers, to the
observer, is only slight. The measures for $m_{1}$ and $m_{2}$ can
be directly bit-counted by independently compressing each continuer.
But assuming we already have the observation-basis expansion $\left|\psi_{m}\right\rangle =a_{1}\left|\psi_{1}\right\rangle +a_{2}\left|\psi_{2}\right\rangle $,
is there no way to compute the probabilities from \emph{this }representation,
simply in terms of $a_{1}\left|\psi_{1}\right\rangle $ and $a_{2}\left|\psi_{2}\right\rangle $,
without counting bits or performing optimal compressions? Considering
that $\mathcal{C}\left(m_{1}\right)$ is already almost identical
to $\mathcal{C}\left(m\right)$, clearly $a_{1}$ must be very high.
This follows straightforwardly from the definition of the DFT. A \emph{low}
$\triangle H_{i}$ alteration-bit-count means a \emph{large} amplitude
in the preferred expansion. By the same reasoning, a \emph{high }alteration-bit-count
corresponds to a \emph{small }amplitude, and since $\mathcal{C}\left(m_{2}\right)$
contributes almost nothing to $\mathcal{C}\left(m\right)$, $a_{2}$
will be tiny.

It thus makes sense for the alteration bits (or change in entropy)
to be reflected in the observation-basis expansion as norm-amplitudes.
It also makes sense that the formula for a norm-amplitude-based measure
will be norm-amplitude-squared, since this is a conserved measure
across unitary transformations that obeys additivity---and due to
Gleason-Busch, we know it is the \emph{only} such measure. And under
this alternative view of probability, the idea of probability interference
is no longer mysterious, as it is an inherent and expected feature
of the DFT frequency-based representation scheme.

Returning to the replicator example, simply replace the three replicator
copies with three different programs, and we end up with the same
basic justification for our new algorithmic measure, just applied
to the domain of programs rather than physical subsystems.

Some readers may fear I have artificially chosen an optimal compression
scheme that just happens to be the discrete form of the solved Schr{\"o}dinger
equation. However, keep in mind that we are permitted in this exercise
to assume the wavefunction under wavefunction realism. I would, however,
challenge anyone concerned about the choice of the DFT to propose
an alternative \emph{a priori }guess for an optimal compression scheme
that they consider to be at least as good, and our discussion can
proceed from there. Considering that, in terms of real world practical
technology, the DFT or something very like it, is at the heart of
nearly all practical synthetic (lossy) compression schemes, it seems
a very plausible choice. 

It is worth noting, in addition, that while real-world lossy compression
algorithms can take more specific forms than the relatively straightforward
DFT compression used in this paper, the resulting algorithms are still
usually describable as ``transform compression'' or ``modified
DFT'' algorithms, being based on essentially the same scheme. Given
that branch-dependence is a general property of algorithmic continuers,
independent of which compression algorithm is used, basis-independence
should arise given just about any transform-based compression algorithm,
and the Born rule will still follow.

\bibliographystyle{eptcs}

\end{document}